\documentclass[aps,prl,showpacs,twocolumn]{revtex4}
\usepackage{amssymb}
\usepackage{epsfig}
\usepackage{graphicx}
\usepackage{amsmath}
\usepackage{array,color}

\begin{document}

\title{Optomechanically-Based Probing of Spin-Charge
Separation in Ultracold Gases}
\author{Qing Sun$^{1}$,  W. M. Liu$^1$,
and An-Chun Ji$^{2,1}$}
\address{$^1$Beijing
National Laboratory for Condensed Matter Physics, Institute of
Physics, Chinese Academy of Sciences, Beijing 100190, China}
\address{$^2$Department of Physics,
Capital Normal University, Beijing 100048, China}
\date{{\small \today}}

\begin{abstract}
We propose a new approach to investigate the spin-charge separation
in 1D quantum liquids via the optomechanical coupled atom-cavity
system. We show that, one can realize an effective two-modes
optomechanical model with the spin/charge modes playing the role of
mechanical resonators. By tuning the weak probe laser under a pump
field, the signal of spin-charge separation could be probed
explicitly in the sideband regime  via cavity transmissions.
Moreover, the spin /charge modes can be addressed separately by
designing the probe field configurations, which may be beneficial
for future studies of the atom-cavity systems and quantum many-body
physics.

\end{abstract}
\pacs{37.30.+i, 03.75.Kk, 42.50.Pq} \maketitle

One dimensional quantum liquids have been fascinating for condensed
matter physicists for quite a few decades \cite{Giamarchi}. For a 1D
quantum liquid, the low energy behavior of the system lies in a
universal class \cite{Haldane} which results in a  remarkable
phenomenon that a single-particle excitation would fractionizes into
a collective charge and spin parts and separates. However, the clear
observation of this phenomenon has proven to be challenging in the
past decades in solid state materials \cite{Jompol}. Recently, the
low dimension quantum fluids have been successfully realized in cold
atom context \cite{Gorlitz}. The unprecedented tunability of
interaction and dimensionality make it a powerful tool to explore
Luttinger liquid \cite{Moritz} or Tonks gas \cite{Paredes,Kinoshita}
in 1D. Stimulated by these experimental advances, some authors
propose to detect spin-charge separation  in ultracold gases by
tracking a wave-packet motion or analyzing the spectrum of a
single-particle excitation \cite{Recati,Kollath,Kecke,Kleine}.
However, because of the small available spatial and limited time
scales, the explicit signals of the spin-charge separation have not
been observed in cold atom experiments so far.

Very recently, cavity optomechanics with cold atoms
\cite{Gupta,Murch} or a BEC \cite{Brennecke2} has acquired
remarkable achievements. In such experiments,  the low energy
collective excitation of cold atoms behaves as a ``moving mirror''
\cite{Sun1,Kanamoto}, which can be detected conveniently by cavity
transmissions. And therefore, this offers a unique method to probe
the low energy excitations of particular quantum phases in ultracold
gases \cite{Sun1}. Based on these advances, we prose a new procedure
in this Letter to detect spin-charge separation in 1D quantum
liquids by considering a 1D two-component ultracold fermionic gases
coupled to a polarization-degenerate optical cavity with external
pump and probe fields. We show that, by tuning the weak probe field,
the explicit signal of spin-charge separation could be probed
definitely via transmission spectra within current experimental
setups. This technique, which has the advantage of nondemolition
measurements \cite{Mekhov} and involves no added complications,
provides a new
practical way to explore the quantum many-body physics in future.
\begin{figure}[t]
\centering
\includegraphics[width=0.4\textwidth]{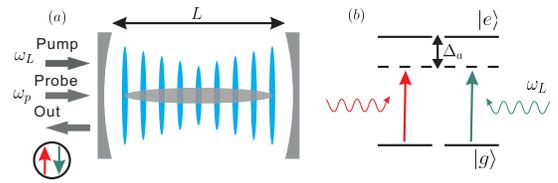}
\caption{(color online). (a) A collection of two-component hyperfine
fermionic atoms are confined in an effectively 1D trap inside an
optical cavity of length $L$. The cavity mode is driven by a ``Pump"
laser of frequency $\omega_L$, and a weak ``Probe" laser of
frequency $\omega_p$ is added to stimulate the system's optical
response. ``Out" is the transmission field. Both the fields are
polarization dependent. (b) Internal energy levels of the
two-component atoms with detuning $\Delta_a=\omega_L-\omega_a$.
Here, $\omega_a$ is the resonant frequency between the ground
$|g\rangle$ and excited $|e\rangle$ states.}
 \label{scheme}
\end{figure}

The system under investigation is illustrated in Fig.
\ref{scheme}(a), where two-component hyperfine fermionic atoms  of
mass $M$ with resonant frequency $\omega_a$ are confined in a 1D
trap inside an optical cavity. The cavity mode of frequency
$\omega_0$ is driven by a pump laser, and we also add a weak probe
field to the cavity, which behaves as a small perturbation to
stimulate the fluctuations of the system \cite {Schliesser}. To
explore the spin-charge separation, both the pump and the probe
fields are polarization dependent. The cavity field couples to the
atomic internal state (see Fig. \ref{scheme}(b)) and induces a
quantized potential on atoms in the far-off resonance limit. Then,
in the dipole and rotating-wave approximations, the atomic part of
Hamiltonian can be written as \cite{Maschler}
\begin{eqnarray}
\hat{H}_a\!&=&\!\sum_{\sigma}\int dx\hat{\Psi}_{\sigma}^{\dag}(x)
[\frac{\hat{P}^2_{x}}{2M}+\hbar U^{\sigma}_{0}\cos^2(Kx)
\hat{c}_{\sigma}^{\dag}\hat{c}_{\sigma}]\hat{\Psi}_{\sigma}(x)\nonumber\\
&+&g_{1D}\int
dx\hat{\Psi}_{\uparrow}^{\dag}(x)\hat{\Psi}_{\uparrow}(x)
\hat{\Psi}_{\downarrow}^{\dag}(x)\hat{\Psi}_{\downarrow}(x).
\label{Hamiltonian1}
\end{eqnarray}
Here, $\hat{\Psi}_{\sigma}(x)$, $\sigma=\uparrow,\downarrow$ is the
pseudo-spin atomic field operator for two hyperfine fermionic atoms,
$\hat{c}_{\sigma}$ is the cavity filed operator for up/down
polarization, and $U^{\sigma}_{0}=U_{0}=g^2_{0}/\Delta_a$ is the
optical dipole potential strength for a single intracavity photon
with  $K=2\pi/\lambda_c$ the wave-vector of the cavity mode.
$g_{1D}=\frac{4\pi\hbar a_s}{M}$ is the strength of contact
interaction between fermions with opposite spin, and $a_s$ is the
effective 1D  low-energy s-wave scattering length, which can be
tuned by Feshbach resonance.

First, following the standard procedure, we transform the atomic
field operator into momentum representation by
$\hat{\Psi}_{\sigma}(x)=L^{-1/2}\Sigma_{k}\hat{f}_{k,\sigma}e^{ikx}$,
where $\hat{f}_{k,\sigma}$ is the fermion annihilation operator for
a plane wave with wave-vector $k$. Then, Hamiltonian
(\ref{Hamiltonian1}) can be rewritten as
\begin{eqnarray}
\hat{H}\!&=&\!\sum_{k,\sigma}\epsilon(k)\hat{f}^{\dag}_{k,\sigma}
\hat{f}_{k,\sigma} +g_{1D}\sum_{k_1,k_2,q}\hat{f}_{k_1+q,\uparrow}
^{\dag}\hat{f}_{k_1,\uparrow}
\hat{f}_{k_2-q,\downarrow}^{\dag}\hat{f}_{k_2,\downarrow}\nonumber\\
\!&+&\!\sum_{\sigma}\hat{c}^{\dag}_{\sigma}\hat{c}_{\sigma}
[\hbar\Delta_{\sigma}+\frac{1}{4}\hbar
U_0\sum_{k}(\hat{f}^{\dag}_{k+2K,\sigma}\hat{f}_{k,\sigma}+{\rm
h.c.})], \label{Hamiltonian2}
\end{eqnarray}
where $\epsilon(k)=\hbar^{2}k^2/2M$ is the single particle kinetic
energy and $\Delta_{\sigma}=\omega_{0}-\omega_{L}+U_{0}N_{\sigma}/2$
is the effective cavity detuning.  Here, we concern the
spin-balanced case with $N_{\sigma}=N$ and $\Delta_{\sigma}=\Delta$.

We shall work in the low photon numbers limit and consider only the
lowest momentum transfer of $2K$ induced by photons. For low
temperature and small momentum $K\ll k_{F}=\pi N/L$, the
particle-hole excitations  occur around the Fermi surface (Fermi
points in 1D). One may then implement the bosonization procedure
\cite{Giamarchi} by introducing the following bosonic operators
\begin{eqnarray}
\hat{b}^{\nu}_{k,\sigma}=\sqrt{\frac{2\pi}{Lk}}\hat{\rho}_{\sigma}^{\nu}(-k),\>
\hat{b}^{\nu\dag}_{k,\sigma}=\sqrt{\frac{2\pi}{Lk}}\hat{\rho}_{\sigma}^{\nu}(k)\
(k>0). \label{bosonization1}
\end{eqnarray}
Here,
$\hat{\rho}^{\nu}_{\sigma}(k)=\sum_{q}\hat{f}^{\nu\dag}_{k+q,\sigma}
\hat{f}^{\nu}_{q,\sigma}$ are density operators for the right and
left moving fermions with $\nu=R,L$. By further introducing the
charge and spin density bosonic operators  $\hat{b}^{\nu}_{k,
\lambda}=\frac{1}{\sqrt{2}}(\hat{b}^{\nu}_{k,\uparrow}
\pm\hat{b}^{\nu}_{k,\downarrow})$, $\lambda=c,s$, and performing the
Bogoliubov transformations
$\hat{d}^{R}_{k,\lambda}=\cosh\gamma_{\lambda}\hat{b}^{R}_{k,\lambda}
+\sinh\gamma_{\lambda}\hat{b}^{L\dag}_{k,\lambda}$,
$\hat{d}^{L\dag}_{k,\lambda}=\sinh\gamma_{\lambda}\hat{b}^{R}_{k,\lambda}
+\cosh\gamma_{\lambda}\hat{b}^{L\dag}_{k,\lambda}$ with
$\tanh2\gamma_\lambda=\frac{\pm g_{1d}}{2\pi v_F\pm g_{1d}}$; we
derive the effective optomechanical model of the coupled system
\begin{eqnarray}
\hat{H}_{\rm{eff}}&=&\sum_{\nu,\lambda}\hbar\omega_{q,\lambda}
\hat{d}^{\nu\dag}_{q,\lambda
}\hat{d}^{\nu}_{q,\lambda}+\sum_{\nu,\lambda}\hbar\widetilde{U}_{\lambda}
\hat{n}_{\lambda}(\hat{d}^{\nu\dag}_{q,\lambda}
+\hat{d}^{\nu}_{q,\lambda})\nonumber\\
&+&\sum_{\sigma}\hbar\Delta\hat{n}_{\sigma}, \label{eff}
\end{eqnarray}
where the first term describes the charge-spin fluctuations of the
1D interacting gas, which play the role of mechanical resonators
with frequency $\omega_{q\equiv2K,\lambda}=2Ku_{\lambda}$. Here,
$u_c=v_F\sqrt{(1+\frac{g_{1D}}{2\pi v_F})^2-(\frac{g_{1D}}{2\pi
v_F})^2}$ and $u_s=v_F\sqrt{(1-\frac{g_{1D}}{2\pi
v_F})^2-(-\frac{g_{1D}}{2\pi v_F})^2}$ are the sound velocities of
the charge-spin excitations for $g_{1D}/\pi v_F\ll1$. The second
term is the coupling between the mechanical modes and cavity fields
with $\widetilde{U}_{\lambda}=\frac{U_{0}}{4}\sqrt{\frac{KL}{\pi}}
(\cosh\gamma_{\lambda}-\sinh\gamma_{\lambda})$ and $
\hat{n}_{c,s}=\hat{n}_{\uparrow}\pm\hat{n}_{\downarrow}$.
\begin{figure}[t]
\centering
\includegraphics[width=0.5\textwidth]{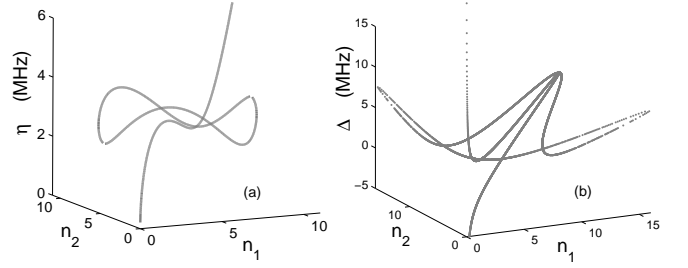}
\caption{Steady-state behavior  of the effective optomechanical
model of coupled  spin-charge excitation modes. (a)-(b) show the
mean-field intracavity up/down polarized photon number $n_1/n_2$
versus the pump rate $\eta$ for $\Delta=(\omega_c+\omega_s)/2$, and
versus detuning $\Delta$ for $\eta/\kappa=4$.}
\label{steadysolution}
\end{figure}

To describe the dynamics of the above driven optomechanical model,
we introduce the quadratures of the mechanical oscillators
$\hat{X}_{\lambda}=\sum_{\nu}\hat{X}^{\nu}_{\lambda}$ ,
$\hat{P}_{\lambda}=\sum_{\nu}\hat{P}^{\nu}_{\lambda}$ with
$\hat{X}^{\nu}_{\lambda}=(\hat{d}^{\nu\dag}_{q,\lambda
}+\hat{d}^{\nu}_{q,\lambda})/\sqrt{2}$,
$\hat{P}^{\nu}_{\lambda}=i(\hat{d}^{\nu\dag}_{q,\lambda
}-\hat{d}^{\nu}_{q,\lambda})/\sqrt{2}$. Then, we arrive at the
coupled Heisenberg-Langevin equations
\begin{eqnarray}
\frac{d\hat{X}_{\lambda}}{dt}\!&=&\!\omega_{\lambda}\hat{P}_{\lambda},
~~~\frac{d\hat{P}_{\lambda}}{dt}=-\omega_{\lambda}
\hat{X}_{\lambda}-2\sqrt{2}
\widetilde{U}_{\lambda}\hat{n}_{\lambda},\nonumber\\
\frac{d\hat{c}_{\sigma}}{dt}\!&=&\!-i[\Delta+\sqrt{2}
(\widetilde{U}_{c}\hat{X}_{c}+\sigma\widetilde{U}_{s}\hat{X}_{s})]
\hat{c}_{\sigma}-\kappa\hat{c}_{\sigma}\!+\!\sqrt{2\theta\kappa}
s^{\sigma}_{\rm{in}},\nonumber\\
\label{Langevin}
\end{eqnarray}
where $\kappa$ is the cavity decay rate and
$s^{\sigma}_{\rm{in}}=\bar{s}_{\sigma}+\delta s_{\sigma}$ denotes
the total amplitude of external fields. Here,
$\bar{s}_{\sigma}\equiv\langle s^{\sigma}_{\rm{in}}\rangle$
represents the pump field, and $\delta s_{\sigma}$ is a small
perturbation, which is induced by the weak probe field with $\delta
s_{\sigma}\equiv s^{\sigma}_p$. $\theta$ is the tunable coupling
parameter.

To proceed, we first consider briefly the  steady-state behavior of
the coupled system. The mean-field solutions of Eqs.
(\ref{Langevin}) are $\bar{P}_{\lambda}=0$,
$\bar{X}_{\lambda}=-2\sqrt{2}\widetilde{U}_{\lambda}
\bar{n}_{\lambda}/\omega_{\lambda}$, and
\begin{eqnarray}
\bar{n}_{\sigma}=\frac{\eta^{2}_{\sigma}}{\kappa^2
+[\Delta-4(\widetilde{U}^2_{c}
\omega^{-1}_{c}\bar{n}_{c}+\sigma\widetilde{U}^2_{s}
\omega^{-1}_{s}\bar{n}_{s})]^2} \label{steady}
\end{eqnarray}
with $\eta_\sigma=\sqrt{2\theta\kappa}\bar{s}_\sigma$. In Fig.
\ref{steadysolution}, we present the mean-field intracavity photon
numbers versus the pump rate and detuning (see below for the
parameters used here). It can be shown that the steady-state driven
by the pump field exhibits optical multi-stability \cite{Sun2}. This
is a characteristic phenomenon of the two-modes optomechanical
system, where both the spin and charge modes are strongly coupled
with the cavity fields.

While in our proposal, we mainly concern the system's optical
response to the weak probe field perturbation in the presence of a
steady state. For a symmetrical pump
$\eta_{\sigma}=\eta_{\bar{\sigma}}$, the steady solution
$\bar{n}_{\sigma}=\bar{n}_{\bar{\sigma}}=\bar{n}$ exists
\cite{Sun3}, which gives rise to
$\bar{n}_{c}=\bar{n}_{\uparrow}+\bar{n}_{\downarrow}=2\bar{n},
~~~\bar{X}_{c}=-4\sqrt{2}\widetilde{U}_{c}\bar{n}/\omega_{c}$ and
$\bar{n}_{s}=\bar{n}_{\uparrow}-\bar{n}_{\downarrow}=0,
~~~\bar{X}_{s}=0$. Then, the optical response to the probe field is
obtained via a linearization of  Eqs. (\ref{Langevin}) around the
steady-state
\begin{eqnarray}
\frac{d\delta\hat{X}_{\lambda}}{dt}&=&
\omega_{\lambda}\delta\hat{P}_{\lambda},
~~~\frac{d\delta\hat{P}_{\lambda}}{dt}
=-\omega_{\lambda}\delta\hat{X}_{\lambda}
-2\sqrt{2}\widetilde{U}_{\lambda}
\sqrt{\bar{n}}\delta\hat{\mathfrak{X}}_{\lambda},
\nonumber\\
\frac{d\delta\hat{\mathfrak{P}}_{\lambda}}{dt}&=&\sqrt{2\kappa}\delta
\mathfrak{P}^{\lambda}_{\rm{in}}-\widetilde{\Delta}\delta
\hat{\mathfrak{X}}_{\lambda}
-\kappa\delta\hat{\mathfrak{P}}_{\lambda}-2\sqrt{2}
\widetilde{U}_{\lambda}\sqrt{\bar{n}}\delta\hat{X}_{\lambda},\nonumber\\
\frac{d\delta\hat{\mathfrak{X}}_{\lambda}}{dt}&=&\sqrt{2\kappa}\delta
\mathfrak{X}^{\lambda}_{\rm{in}}+\widetilde{\Delta}\delta
\hat{\mathfrak{P}}_{\lambda}-\kappa\delta\hat{\mathfrak{X}}_{\lambda},
\label{probe}
\end{eqnarray}
with
$\widetilde{\Delta}=\Delta-8\widetilde{U}^2_{c}\bar{n}\omega^{-1}_c$.
Here, $\delta\hat{\mathfrak{X}}_{c,s}
=(\delta\hat{\mathfrak{X}}_{\uparrow}\pm\delta
\hat{\mathfrak{X}}_{\downarrow})/\sqrt{2}$,
$\delta\hat{\mathfrak{P}}_{c,s}
=(\delta\hat{\mathfrak{P}}_{\uparrow}\pm\delta
\hat{\mathfrak{P}}_{\downarrow})/\sqrt{2}$ represent the
cavity-field charge-spin quadratures with
$\delta\hat{\mathfrak{X}}_{\sigma}=(\delta\hat{c}^{\dag}_{\sigma}
+\delta\hat{c}_{\sigma})/\sqrt{2}$,
$\delta\hat{\mathfrak{P}}_{\sigma}=i(\delta\hat{c}^{\dag}_{\sigma}
-\delta\hat{c}_{\sigma})/\sqrt{2}$. And
$\delta\mathfrak{X}^{c,s}_{\rm{in}}
=(\delta\mathfrak{X}^{\uparrow}_{\rm{in}}\pm\delta
\mathfrak{X}^{\downarrow}_{\rm{in}})/\sqrt{2}$,
$\delta\mathfrak{P}^{c,s}_{\rm{in}}=
(\delta\mathfrak{P}^{\uparrow}_{\rm{in}}\pm\delta
\hat{\mathfrak{P}}^{\downarrow}_{\rm{in}})/\sqrt{2}$ denote the
corresponding probing field terms with
$\delta\mathfrak{X}^{\sigma}_{\rm{in}} =(s^{\sigma*}_{p}
+s^{\sigma}_{p})/\sqrt{2}$, $\delta\mathfrak{P}^{\sigma}_{\rm{in}}
=i(s^{\sigma*}_{p} -s^{\sigma}_{p})/\sqrt{2}$. For cold-atom system,
the damping of the  spin-density excitations are much smaller than
the $\omega_\lambda$, and therefore can be neglected.
\begin{figure}[t]
\centering
\includegraphics[width=0.5\textwidth]{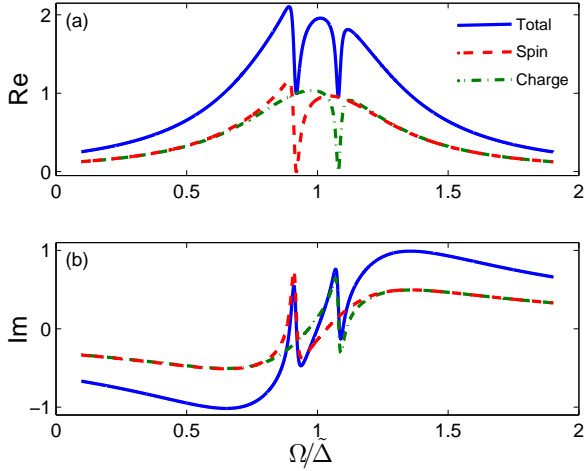}
\caption{(color online.) (a) Real and (b) Image parts of the
intracavity field response versus $\Omega$ (in unit of
$\widetilde{\Delta}$). The solid line shows the  total response
$\mathfrak{R}_{\uparrow (\downarrow)}$  to the polarized probe
field, and the dash-dotted and dashed lines show the responses
$\mathfrak{R}_{c}/\mathfrak{R}_{s}$ to the charge/spin modes
respectively.} \label{quadrature}
\end{figure}

We note that, although both the two mechanical modes are coupled
nonlinearly with cavity field  in Eqs. (\ref{Langevin}), the
fluctuations of the spin and charge modes in the above Eqs.
(\ref{probe}) can be excited independently. This  is a unique
feature of the system, which encodes the explicit signal of
spin-charge separation. To see this, we transform  Eqs.
(\ref{probe}) into frequency space in the rotating frame. Here both
the fluctuations of the mechanical and cavity field variables
oscillate at frequencies $\pm\Omega$ around the steady-state, with
$\Omega=\omega_p-\omega_L$ being the frequency difference between
the probe and pump fields. Then, the intracavity field amplitude can
be derived as \cite{Sun4}
\begin{eqnarray}
A_{\lambda}[\Omega]=\frac{1+if_{\lambda}
(\Omega)}{-i(\Omega-\widetilde{\Delta})+\kappa+2\widetilde{\Delta}
f_{\lambda}(\Omega)}\sqrt{2\theta\kappa}s_p^\lambda,
\label{amplitude}
\end{eqnarray}
with
\begin{eqnarray}
f_{\lambda}(\Omega)=\frac{4\widetilde{U}^2_{\lambda}
\bar{n}\omega_{\lambda}}{\kappa-i(\Omega+\widetilde{\Delta})}
\frac{1}{\Omega^2-\omega^2_{\lambda}}.
\end{eqnarray}
Here, $s^\lambda_p=(s^{\uparrow}_p\pm s^{\downarrow}_p)/2$
represents the input charge/spin probe field amplitudes.

Before proceeding, we consider the following experimental achievable
parameters: $L\sim100$ $\mu$m, $N\simeq 5000$ Alkali metal atoms
({\it e.g.} $^{87}$Rb, $M=1.5\times10^{-25}kg$) and $U_0=20$ KHz,
which give rise to $\omega_\lambda\sim$ several MHz and
$U_\lambda\sim100$ KHz. Then cavity dumping $\kappa$ can be chosen
to satisfy $U_\lambda\sqrt{\bar{n}}\ll\kappa\ll\omega_\lambda$,
which places the system well in the resolved sideband regime
\cite{Marquardt}. In this regime, there exists normal mode splitting,
and  the $-\Omega$ part of cavity fluctuations can be neglected,
which may enable further simplification of the solutions of Eq.
(\ref{amplitude}).
\begin{figure}[t]
\centering
\includegraphics[width=0.5\textwidth]{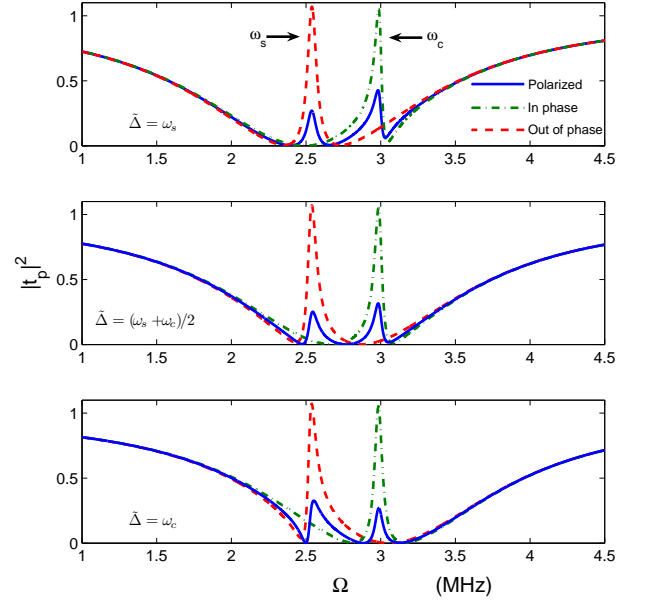}
\caption{(color online.) Transmission spectra for different values
of detuning $\widetilde{\Delta}$ and three kinds of probe fields
with polarized (solid), in-phase (dashed) and out-of-phase
(dash-dotted) configurations.} \label{transmission}
\end{figure}

In experiments, one of the most important observable quantity is the
polarized  intracavity field response to the probe field
$\mathfrak{R}_\sigma[\Omega]\equiv
\sqrt{2\theta\kappa}A_\sigma[\Omega]/s^{\sigma}_p$, which reads
\begin{eqnarray}
\mathfrak{R}_\sigma[\Omega]=\alpha \mathfrak{R}_{c}[\Omega]
+\sigma\beta\mathfrak{R}_{s}[\Omega],
\end{eqnarray}
with $\mathfrak{R}_{\lambda}[\Omega]=\sqrt{2\theta\kappa}
A_\lambda[\Omega]/s^{\lambda}_p$ and $\alpha=s^{c}_p/s^\sigma_p$,
$\beta=s^{s}_p/s^\sigma_p$ determined by the configuration of probe
field. By numerically solving Eq. (\ref{amplitude}) in the sideband
regime, we show the main results of the intracavity field responses
to the polarized probe field in Fig. \ref{quadrature} with
$\alpha=\beta=1/2$ and $\kappa=2\pi\times 150$ KHz for illustration.
A remarkable feature of the spectrum $\mathfrak{R}_{\uparrow
(\downarrow)}$ is that it demonstrates a well defined double dips
centered at the charge/spin modes. The underlying mechanism can be
understood as follows: when we input a small probe field to disturb
the system around the steady state stablized by the pump field, the
corresponding intracavity field response sets up, which is genrally
comprised of the response of charge/spin modes. When the frequency
difference between the probe and pump fields $\Omega$ is tuned to be
one of the mode frequencies, the corresponding collective mode would
be excited resonantly, and accordingly the cavity response would
drop dramatically. Therefore in this scheme, one doesn't have to
track the motions of spin/charge wave-packets \cite{Recati}, which
have different velocities; we only need to detect the resonant
frequencies of the collective spin/charge excitations by tuning the
probe fields.

\begin{figure}[t]
\centering
\includegraphics[width=0.35\textwidth]{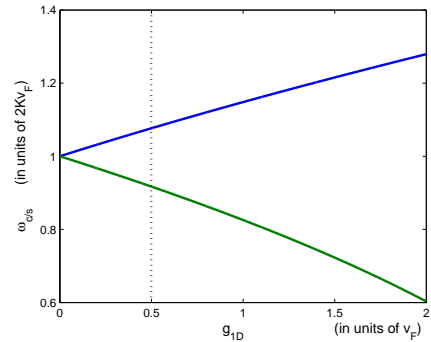}
\caption{Frequencies of the charge ($\omega_c$) and spin
($\omega_s$) modes versus interacting parameter $g_{1D}$ for giving
$K$. The dotted line marks the parameter used in  Fig.
\ref{quadrature} and Fig. \ref{transmission}.} \label{frequency}
\end{figure}

To further clarify, we present the probe power transmission
$|t^{\sigma}_p|^2=|1-\mathfrak{R}_\sigma|^2$ for different probe
field configurations  (here, the critical pump $\theta=1/2$ is
assumed) in Fig. \ref{transmission}, where the charge/spin modes are
also clearly resolved (solid line). We see that, by adjusting the
probe field configuration to be  in-phase (dashed line) or
out-of-phase (dash-dotted line), the charge/spin modes can be
addressed separately. In Fig. \ref{transmission}, We also
investigate the impact of detuning $\widetilde{\Delta}$ on the
spectra. It can be seen that although the transmission is generally
modified, the peaks of the probe spectra always occur at the
charge/spin modes, which are independent of detuning
$\widetilde{\Delta}$. The frequencies of the charge and spin modes
versus interacting parameter $g_{1D}$ for $g_{1D}/\pi v_F<1$ are
shown in Fig. \ref{frequency}, which could also be inspected in
future experiments.

Until now, we have  mainly focused on the femionic gas.
Experimentally, the above scheme could also be realized in
two-component Bose gas by implementing the hydrodynamical theory
\cite{Sun1}. The velocities of the charge/spin excitations are
$u_{c,s}=u_0\sqrt{1\pm g_{12}/g}$ with $g$ and $g_{12}$ the
intraspecies and interspecies interactions \cite{Kleine}. The
effective frequencies of the collective modes are of order MHz in
Luttinger liquid regime and lies well in  the resolved sideband
limit. Further studies will consider trapping potentials, where the
frequency of collective modes have to be time-averaged in a period
because of the position-dependent velocities of the excitations
\cite{Recati}.

In summary, we have shown how to implement the probing of the  the
intriguing spin-charge separation in 1D quantum liquids via the
optomechanical coupled atom-cavity system. Such experiments allow us
to determine definitely the collective excitations of the 1D
strongly correlated system with nondemolition measurements.
Furthermore, the two-modes optomechanics itself may be of interests
for future studies of quantum physics.

We acknowledge T. Esslinger for private communications and thank Hui
Hu for helpful discussions. This work is supported by NSFC under
grants No. 11704175, the NKBRSFC under grants No. 2011CB921500.


\begin{thebibliography}{99}
\bibitem{Sun3} The solution with $\bar{n}_{\sigma}\neq\bar{n}_{\bar{\sigma}}$
may also exists (see Fig. \ref{steadysolution}). Here, we consider
the stable branch below the threshold of bistability.
\bibitem{Sun4} We note that, because the frequency of the probe field is
$\omega_p=\omega_L+\Omega$, the transmission depends only on the
$+\Omega$ part of the cavity field amplitudes.
\end{thebibliography}
\end{document}